\documentclass[11pt]{openjournal}

\usepackage{amsmath,amssymb,amsfonts}
\usepackage{bm}
\usepackage{graphicx,float,xcolor}
\usepackage[pdftex, colorlinks=true, linkcolor=blue, urlcolor=blue,  citecolor=blue]{hyperref}

\shorttitle{Galaxy clustering growth forecasts with multipoles}

\begin{document}

\title{Assessing non-linear models for galaxy clustering I: \\
        unbiased growth forecasts from multipole expansion}

\author{Katarina Markovi\v{c},}
\address{Institute of Cosmology \& Gravitation, University of Portsmouth,
Dennis Sciama Building, Burnaby Road, Portsmouth PO1 3FX, UK}
\email{dida.markovic@port.ac.uk}

\author{Benjamin Bose,}
\address{Departement de Physique Theorique, Universite de Geneve, 24 quai Ernest Ansermet, 1211 Geneve 4, Switzerland}
\email{benjamin.bose@unige.ch}

\and

\author{Alkistis Pourtsidou}
\address{School of Physics and Astronomy, Queen Mary University of London, Mile End Road, London E1 4NS, UK}
\address{Department of Physics \& Astronomy, University of the Western Cape, Cape Town 7535, South Africa}
\email{a.pourtsidou@qmul.ac.uk}

\begin{abstract}
\noindent
We assess the performance of the Taruya, Nishimichi and Saito (TNS) model for the halo redshift space power spectrum, focusing on utilising mildly non-linear scales to constrain the growth rate of structure $f$. Using simulations with volume and number density typical of forthcoming Stage IV galaxy surveys, we determine ranges of validity for the model at redshifts $z=0.5$ and $z=1$. We proceed to perform a Bayesian MCMC analysis utilising the monopole, quadrupole, and hexadecapole spectra, followed by an exploratory Fisher matrix analysis. As previously noted in other forecasts as well as in real data analyses, we find that including the hexadecapole can significantly improve the constraints. However, a restricted range of scales is required for the hexadecapole in order for the growth parameter estimation to remain unbiased, limiting the improvement. We consistently quantify these effects  by employing the multipole expansion formalism in both our Fisher and MCMC forecasts.
\end{abstract}

\maketitle


\section{Introduction}
Forthcoming large-scale structure (LSS) surveys such as EUCLID\footnote{\url{www.euclid-ec.org}} \citep{Laureijs:2011gra}, WFIRST\footnote{\url{https://wfirst.gsfc.nasa.gov/}}, and the Dark Energy Spectroscopic Instrument (DESI)\footnote{\url{www.desi.lbl.gov}}, are promising to deliver exquisite cosmological measurements and test
the laws of gravity with unprecedented precision. The standard cosmological model, $\Lambda$CDM, has been shown to provide an excellent fit to the data from a suite of CMB and LSS surveys \citep{Planck:2015xua,Anderson:2012sa,Song:2015oza,Beutler:2016arn}, assuming that General Relativity is the correct description of gravity across all scales. A plethora of modified gravity models have been proposed, motivated by the possibility of explaining the late time accelerated expansion of the Universe without the need of a cosmological constant \citep[see][for a comprehensive review]{Clifton:2011jh}.
Future LSS measurements with Stage IV spectroscopic galaxy surveys are aiming to measure the \emph{logarithmic growth rate of structure} $f$, which strongly depends on cosmology and gravity \citep{Guzzo:2008ac}. 
This will be achieved by probing the redshift space distortion (RSD) signature with the galaxy power spectrum or correlation function \citep{Blake:2011rj,Reid:2012sw,Macaulay:2013swa,Beutler:2013yhm,Gil-Marin:2015sqa,Simpson:2015yfa}.
\newline
\newline
The volume and number of galaxies probed by Stage IV surveys are very large, meaning that these surveys will not be limited by statistical uncertainties, but rather by our ability to deal with systematic effects and theoretical uncertainties. The latter include non-linear effects on the redshift space galaxy clustering. In the case of RSD measurements, these have to be understood and modelled properly so that we can confidently utilise the data from mildly non-linear scales to improve the constraints on $f$. 
\newline
\newline
A way to attack this challenge is by using perturbation theory based models \citep{Bernardeau:2001qr,Kaiser:1987qv,Scoccimarro:2004tg} that can be easily extended to include non-standard theories of gravity and dark energy \citep[see][for example]{Bose:2016qun,Bose:2017dtl,Bose:2017jjx,Bose:2018orj}. These can be combined with phenomenological ingredients to model \emph{non-linear physics} \citep{Taruya:2010mx,Senatore:2014vja,delaBella:2017qjy}. Additionally, a model for the \emph{galaxy bias} is required to relate the dark matter and galaxy distributions . 
\newline
\newline
An important factor, which is the main focus of this paper, is our ability to use these prescriptions to model non-linear structure formation in an unbiased manner, and how this affects the growth of structure parameter estimation with Stage IV surveys.
\newline
\newline
For this purpose, we consider the commonly used TNS model \citep{Taruya:2010mx}. This model can reproduce the broadband power spectrum including RSD from simulations at linear and mildly non-linear scales \citep{Nishimichi:2011jm,Taruya:2013my,Ishikawa:2013aea,Zheng:2016zxc,Gil-Marin:2015sqa,Gil-Marin:2015nqa,Bose:2017myh,Bose:2016qun}. 
The model is combined with the generalised bias prescription of \citet{McDonald:2009dh}. This is very similar to what has been used in part of the BOSS survey data analysis \citep{Beutler:2016arn}.
We use a set of COLA simulations \citep{Tassev:2013pn,Howlett:2015hfa, Valogiannis:2016ane,Winther:2017jof} to determine a range of validity for the models at redshifts $z=0.5$ and $1$. We then perform a Bayesian MCMC analysis  with the goal of assessing the constraining power of Stage IV-like surveys while also confirming the validity of the models across the range of scales we determined previously. Our focus is getting unbiased constraints on $f$ using information from the first three multipoles: the monopole ($P_0$), quadrupole ($P_2$), and hexadecapole ($P_4$) of the power spectrum.  
We then move on to an exploratory Fisher matrix forecast analysis using the full anisotropic power spectrum, $P(k,\mu)$, in addition to the multipole expansion formalism. Our final estimates for the unbiased measurements on $f$ are determined consistently from the MCMC and Fisher multipole expansion forecasts analyses.
\newline
\newline  
This paper is organised as follows: In \autoref{sec:model} we present the TNS-based biased tracer RSD model we assemble, and the fits to simulations to determine the fiducial nuisance parameters and model range of validity. In \autoref{sec:mcmc} we perform a Bayesian MCMC analysis and present results, followed by a Fisher matrix analysis in \autoref{sec:fisher}. We then present a comparison between Fisher and MCMC using the multipole expansion formalism. In \autoref{sec:summary} we summarise our findings and conclude. Our Fisher matrix codes have been made publicly available at \url{https://github.com/Alkistis/GC-Fish-nonlinear}.

\section{TNS model description and comparison to simulations}
\label{sec:model}

\subsection{Model description}

The TNS model \citep{Taruya:2010mx} is based on standard Eulerian perturbation theory (SPT), which assumes that the background space-time is homogeneous and isotropic, and that we work within the \emph{Newtonian regime} at mildly non-linear scales. These scales are far within the horizon but with $\delta, \theta \ll 1$, where $\delta$ and $\theta$ are the density and velocity perturbations respectively. We also assume that gravity is described by general relativity. The explicit model is given by 
 \begin{align}
 P^{\rm S}_{\rm (TNS)}(k,\mu) =& D_{\rm FoG}(\mu^2 k^2 \sigma_v^2)\Big[ P_{g,\delta \delta} (k,b_1,b_2,N) + 2 \mu^2 P_{g,\delta \theta}(k,b_1,b_2) +  \mu^4 P_{\theta \theta} (k) \nonumber \\ & \qquad \qquad \qquad \qquad + b_1^3A(k,\mu) + b_1^4B(k,\mu) +b_1^2 C(k,\mu)  \Big], 
 \label{redshiftps}
 \end{align} 
where $f$ is the logarithmic growth rate, $\mu$ is the cosine of the angle between $\mbox{\boldmath$k$}$ and the line of sight and $P_g$ are the 1-loop galaxy power spectra with the bias model of \citet{McDonald:2009dh} implicitly included \footnote{We assume the local Lagrangian assumption and hence only first and second order bias are considered: $b_1$ and $b_2$ as well as a stochasticity term $N$.}, while the $A$, $B$ and $C$ terms are non-linear perturbative corrections arising from the transformation to redshift space. The terms in brackets are all constructed within SPT\footnote{We refer the reader to \citet{Beutler:2016arn} and \citet{Taruya:2010mx} for explicit expressions for the galaxy spectra and correction terms.}, while the prefactor, $D_{\rm FoG}$, is phenomenological. Here we choose a Lorentzian form \citep{GilMarin:2012nb,Percival:2008sh,Taruya:2010mx,Peacock:1993xg} 
\begin{equation}
    D_{\rm FoG}^{\rm Lor}(k^2\mu^2 \sigma_v^2) = \frac{1}{1 + (k^2\mu^2 \sigma_v^2)/2} \, ,
    \label{dampf}
\end{equation}
where $\sigma_v$ is a redshift-dependent free parameter and represents the velocity dispersion of perturbations at cluster scales. We again refer the reader to \citet{Beutler:2016arn,Taruya:2010mx} for the formulas for the perturbative components of the model, along with the explicit dependency on the independent free bias parameters $\{b_1,b_2,N\}$. As was done in \citet{Beutler:2016arn}, we ignore all other bias terms under the local Lagrangian assumption \citep{Sheth:2012fc,Chan:2012jj,Saito:2014qha,Baldauf:2012hs}.
\newline
\newline 
We emphasise again that this model is very similar that used in the recent BOSS analysis \citep{Beutler:2016arn}. We choose it to make use of its thorough validation with simulations and mock catalogues. It is also  worth noting that it has shown robustness when considering alternative theories of gravity \citep[for example][]{Bose:2016qun}. However, we stress that there are key differences in our model. We choose a Lorentzian rather than a Gaussian damping factor in \autoref{dampf} \footnote{This functional form was shown to give a significantly better fit to COLA multipoles in \citet{Bose:2019psj}.}, we include the $C(k,\mu)$ term in \autoref{redshiftps} and use SPT rather than RegPT \citep{Taruya:2012ut}, the latter being the biggest difference. SPT is known to suffer from divergences in the loop expansion at low redshifts \citep[see][for example]{Carlson:2009it}, which the re-normalisation scheme of the RegPT approach addresses. Despite this, SPT clearly does well for $z\geq 1$ \citep{Carlson:2009it,Osato:2018ldv}. Our work finds that it also works well for $z=0.5$ given our derived goodness of fit to the data. This was also suggested in other works \citep{Carlson:2009it,Vlah:2014nta,Bose:2017dtl}. Further, in \citet{Bose:2019psj} the authors also show that SPT provides a better fit to the redshift space halo multipoles at both $z=0.5$ and $z=1$ than when using the RegPT prescription. Given these findings, it would be very interesting to repeat the BOSS data analysis with this variant of the TNS model.
\newline
\newline
The full set of nuisance parameters in this model is therefore $\{ \sigma_v, b_1, b_2, N \}$. In our MCMC and Fisher matrix analyses we will vary these nuisance parameters along with the cosmological parameter of interest, the growth of structure $f$. Our chosen set of parameters is restricted. Perhaps most importantly, we do not include variations of the background ``shape'' parameters or the Alcock-Paczynski effect. There are two reasons for this. 
Firstly, our  goal is to demonstrate the trade-off between our constraining power on $f$ and the bias on its estimation as a function of the $k$-ranges used from the monopole, quadrupole, and hexadecapole spectra. We then wish to provide a consistent comparison between Fisher matrix and MCMC forecasts. This can be achieved without an extensive set of parameters. 
Secondly, varying extra parameters in the MCMC is computationally expensive, since all the model components have to be calculated for every sample. Using a restricted set mitigates this problem. However, optimising our code for speed is under development and we hope to present an analysis with all parameters of interest in future work. We also note that while it is customary to present constraints on $(f\sigma_8)(z)$, and indeed the BOSS analysis uses this parametrisation, other analyses have opted to present constraints on $f$ alone, e.g. \citet{Blake:2011rj}. The reason for this choice in a real data analysis is to test if the Planck best-fit model also predicts the observed growth of structure by the galaxy survey.
\\ \\
Finally, we note that great progress has been made since the TNS model was first proposed, in particular in relation to the relationship between bias and redshift space distortions (see \citet{Desjacques:2016bnm} for a review). Despite this, the TNS model with this minimal bias model remains one of the simplest and most economical phenomenological models. For example, a fully consistent treatment of the halo power spectrum in redshift space discussed  requires many more free parameters \citep{Desjacques:2018pfv}. 

\subsection{Comparison to Simulations}
\begin{figure}[t!]
\centering
  \includegraphics[scale=0.4]{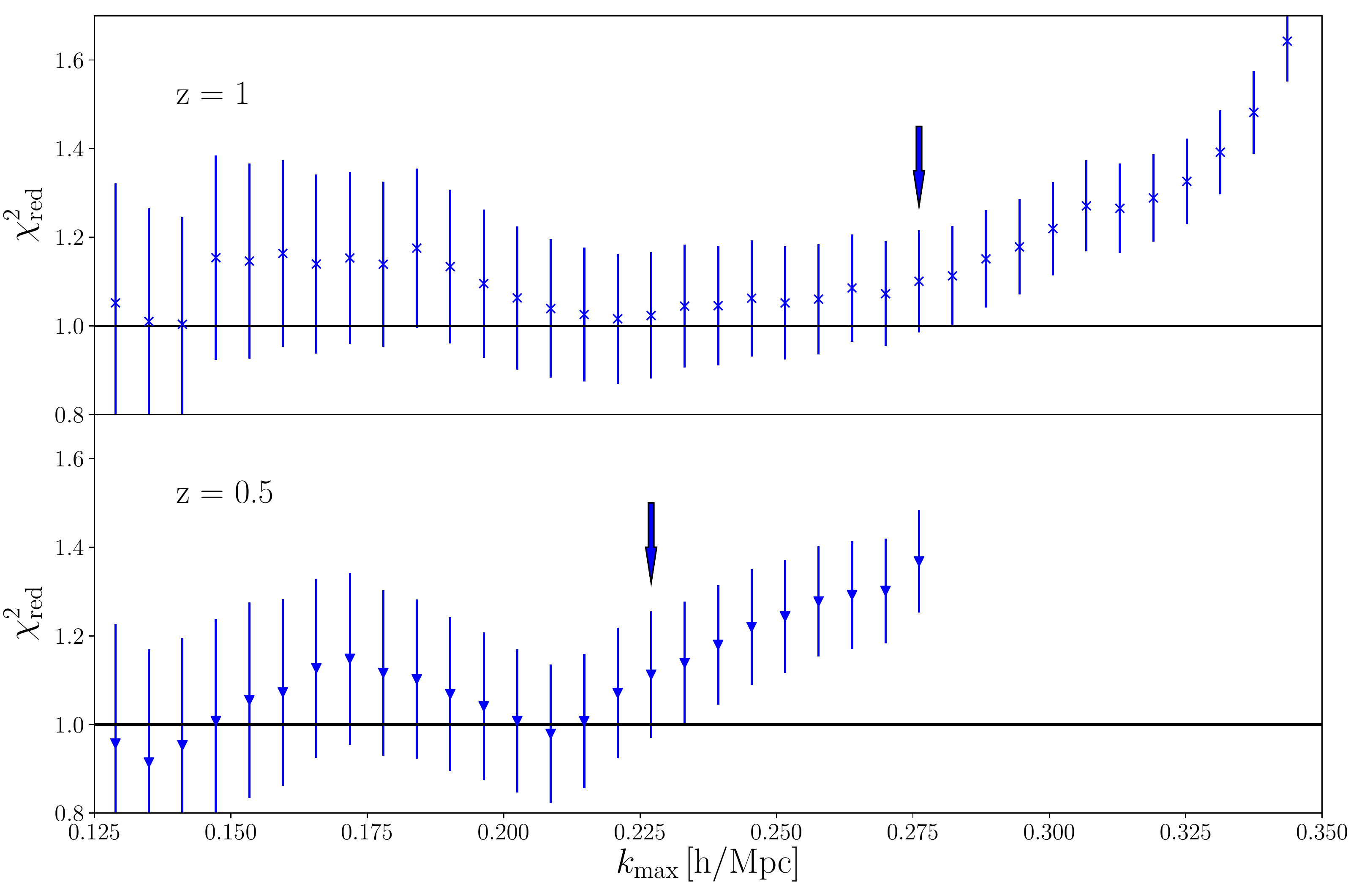}\\
  \caption[CONVERGENCE]{The minimised $\chi_{\rm red}^2$ statistic as a function of $k_{\rm max}$ at $z=1$ (top) and $z=0.5$ (bottom) for the  TNS model. The error bars shown are the $2\sigma$ confidence interval for the $\chi_{\rm red}^2$ statistic with $N_{\rm dof}$ degrees of freedom. The arrows indicate the $k_{\rm max}$ value we use in our fits.}
\label{redc}
\end{figure}

\begin{figure}[t!]
\centering
  \includegraphics[scale=0.4]{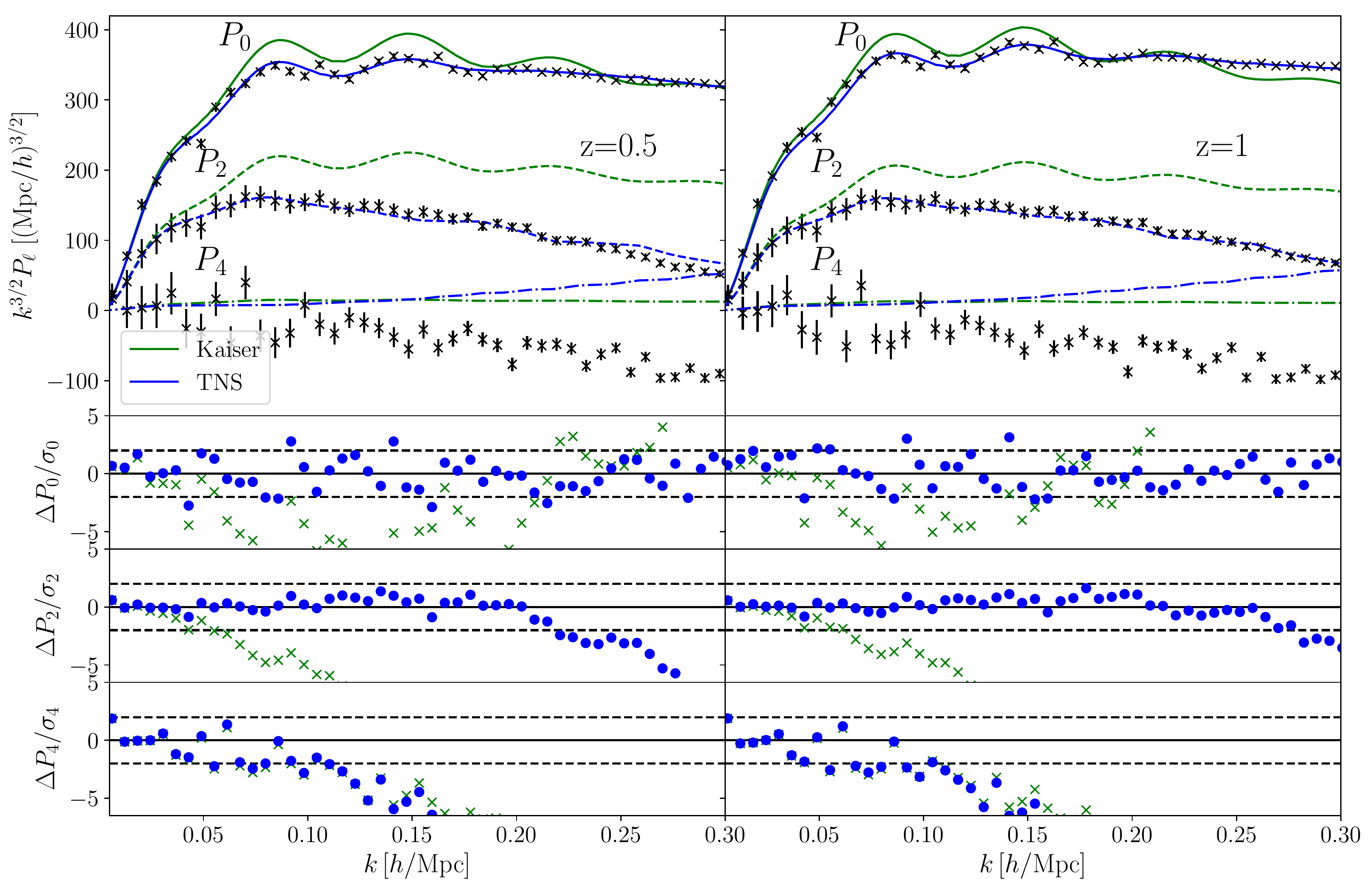} 
  \caption[CONVERGENCE]{The averaged halo monopole, quadrupole and hexadecapole of 4 PICOLA simulations (black points) with errors given by linear theory assuming a survey volume of $V=4$\,~Gpc${}^3/h^3$ and a number density of $n_h = 1\times 10^{-3} \, h^3/\mbox{Mpc}^3$. The best fitting  TNS model is shown as blue curves. The lower panels show the residuals with the data. The dashed lines indicate the $2\sigma$ region around the data. We have also plotted linear theory marked in green for reference.}
\label{fits}
\end{figure}

This section is dedicated to determining a rough range of validity for \autoref{redshiftps} as well as fiducial values for the nuisance parameters. To do this, we make use of a set of four Parallel COmoving Lagrangian Acceleration (PICOLA) simulations \citep{Howlett:2015hfa,Winther:2017jof} of box length $1024 \, \mbox{Mpc}/h$ with $1024^3$ dark matter particles and a starting redshift $z_{\rm ini}=49$. These are all run  within the same $\Lambda$CDM cosmology taken from WMAP9 \citep{Hinshaw:2012aka}: $\Omega_m = 0.281$, $\Omega_b=0.046$, $h=0.697$, $n_s=0.971$ and $\sigma_8(z=0) = 0.844$.
\newline
\newline
 Halo catalogues from these simulations are constructed using the Friends-of-Friends algorithm with a linking length of 0.2-times the mean particle separation. The halo spectra are measured at redshifts of $z=0.5$ and $z=1$ and use all halos above a mass of $M_{\rm min} = 4 \times 10^{12} \, M_{\odot}$.  This corresponds to a number density of $n_{\rm h} = 1\times 10^{-3} \, h^3/\mbox{Mpc}^3$ which is similar to that estimated for Stage IV surveys galaxy number density around the redshifts considered \citep{Amendola:2016saw}. 
\newline
\newline 
As is commonly done in real data analyses, the  power spectrum is decomposed into its multipoles. The PICOLA multipoles are measured using the distant-observer approximation \footnote{That is, we assume the observer is located at a distance much greater then the box size ($r\gg 1024 \, \mbox{Mpc}/h$), so we treat all the lines of sight as parallel to the chosen Cartesian axes of the simulation box. Next, we use an appropriate velocity component ($v_x, v_y$ or $v_z$) to disturb the position of a matter particle.} and averaged over three line-of-sight directions. We further average over the four PICOLA simulations. We note here that (PI)COLA is an approximate method and should be validated against full N-body to ensure any comparisons to the measurements are meaningful. In \citet{Izard:2015dja}  comparisons of COLA with full N-body measurements are made, and sufficient agreement is found in the halo monopole and quadrupole. Furthermore, in \citet{Bose:2019ywu}, comparisons of COLA with full N-body are made for the halo monopole, quadrupole and hexadecapole using the same COLA code and simulation specifications used in this work. The authors find the COLA approach to be sufficiently accurate at the scales and redshifts considered here. 
\newline
\newline
On the theoretical side, the multipoles are expressed as 
\begin{equation}
P_\ell^{\rm S}(k)=\frac{2\ell+1}{2}\int^1_{-1}d\mu P^{\rm S}(k,\mu)\mathcal{P}_\ell(\mu),
\end{equation}
where $\mathcal{P}_\ell(\mu)$ denote the Legendre polynomials and $P^{\rm S}(k,\mu)$ is given by \autoref{redshiftps}. For our fitting analysis, we utilise only the monopole ($\ell=0$) and quadrupole ($\ell=2$) since they include most of the clustering information. The hexadecapole is later considered in \autoref{sec:mcmc}, where we perform an MCMC analysis on the PICOLA data. 
\newline
\newline
Our maximum scale in $k$ (the model's range of validity) will be denoted as $k_{\rm max}$. To determine this we follow the following procedure: 
\begin{enumerate}
    \item
    We fix all cosmological parameters including the growth rate $f$ and perform a least $\chi^2$ fit to the PICOLA data by varying the model nuisance parameters\footnote{This is done through a brute force search of parameter space using the code {\tt MG-Copter} \citep{Bose:2016qun}.}. In our $\chi^2$ analysis we use all data bins from $k_{\rm min} = 0.006 \, h/{\rm Mpc}$ to $k_{\rm max}$, with $k_{\rm max}$ being varied in the range $0.125 \, h/{\rm Mpc} \leq k_{\rm max} \leq 0.3 \, h/{\rm Mpc}$.
    \item 
    We take the $95\%$ ($2\sigma$) confidence intervals ($2 \Delta \chi^2_{\rm red}$) on a $\chi^2$ distribution with $N_{\rm dof}$ degrees of freedom. Since $N_{\rm dof}$ is large in our analysis the errors are approximately symmetric. 
    \item 
    The criterion we use to calculate the final $k_{\rm max}$ for the rest of the analysis is the maximum k-value which gives $\left[\chi^2_{\rm red}(k_{\rm max})~-~2\Delta\chi^2_{\rm red}(k_{\rm max})\right]~\leq~1$.
\end{enumerate}
Roughly, this gives an indication of which range of scales the model is reliable at without biasing cosmological parameter estimates at the required $2\sigma$ level\footnote{This procedure is validated in \autoref{sec:mcmc}.}. The reduced $\chi^2$ statistic is given by 
\begin{equation}
\chi^2_{\rm red}(k_{\rm max}) = \frac{1}{N_{\rm dof}}\sum_{k=k_{\rm min}}^{k_{\rm max}} \sum_{\ell,\ell'=0,2} \left[P^{\rm S}_{\ell,{\rm data}}(k)-P^{\rm S}_{\ell,{\rm model}}(k)\right] \mbox{Cov}^{-1}_{\ell,\ell'}(k)\left[P^{\rm S}_{\ell',{\rm data}}(k)-P^{\rm S}_{\ell',{\rm model}}(k)\right],
\label{covarianceeqn}
\end{equation}
 where $\mbox{Cov}_{\ell,\ell'}$ is the Gaussian covariance matrix between the different multipoles and $k_{\rm min} = 0.006 \, h/{\rm Mpc}$. The number of degrees of freedom $N_{\rm dof}$ is given by $N_{\rm dof} = 2\times N_{\rm bins} - N_{\rm params}$, where $N_{\rm bins}$ is the number of $k-$bins used in the summation and $N_{\rm params}$ is the number of free parameters in the theoretical model. Here, $N_{\rm params} = 4$ for the TNS model of \autoref{redshiftps}.  Finally, the bin-width we use is $\Delta k = 0.006 \, h/{\rm Mpc}$. 
\\ \\ 
We use linear theory for the covariance matrix between the multipoles \citep[see Appendix C of][for details]{Taruya:2010mx}. This has been shown to reproduce N-body results up to $k\leq 0.3 \, h/\mbox{Mpc}$ at $z=1$. A linear covariance also seems to work well at $z=0.5$ up to $k\leq 0.2 \, h/\mbox{Mpc}$ at $z=0.5$, as shown very recently in \cite{Sugiyama:2019ike}. We note that for analysing real data from  Stage IV surveys validated analytical approximations will likely be used alongside numerical covariances constructed using mocks.
In the covariance matrix we assume a number density of $n_h= 1\times 10^{-3} \, h^3/\mbox{Mpc}^3$ and a survey volume\footnote{This volume corresponds to a survey with sky area $A_{\rm sky}\simeq 16,000 \, (14,000) \, \mbox{deg}^2$ and bin width $\Delta z =0.2 \, (0.1)$ around $z=0.5 \, (1)$.} of $V_{\rm s}=4 \, \mbox{Gpc}^3/h^3$. 
\newline
\newline
In \autoref{redc} we show the minimized $\chi^2_{\rm red}(k_{\rm max})$ for $z=0.5$ and $z=1$ for the TNS model, with the associated $2\sigma$ error bars. We determine $k_{\rm max} = 0.227 \, h/{\rm Mpc}$ and $k_{\rm max} = 0.276 \, h/{\rm Mpc}$ at $z=0.5$ and $z=1$ respectively. The larger $k_{\rm max}$ at $z=1$ is expected due to less non-linear structure formation at higher redshifts. We summarise the best fit nuisance parameters and details of the fit in \autoref{fittable}. We also plot the best fit TNS multipoles against the PICOLA data in \autoref{fits}. In the bottom panels of \autoref{fits} we show the residuals, that is the difference in theoretical prediction to simulation measurement divided by the errors coming from the covariance matrix. Linear theory \citep{Kaiser:1987qv} is also shown in green as a reference, where we use $b_1$ as measured from the simulations and the fiducial value of $f$.

\begin{table}[h]
\centering
\caption{Number of bins, $k_{\rm max} [h/{\rm Mpc}]$, and fiducial parameters found by a least $\chi^2$ fit to the PICOLA data.}
\begin{tabular}{| c || c | c | c | c | c | c |}
\hline  
 z &  $N_{\rm bins}$ &   $k_{\rm max}$ &  $b_1$ &  $b_2$ & $N$ &   $\sigma_v$ \\ \hline 
 $0.5$ &$36$ & $0.227$& $1.506$ &$0.091$  &$-272$ & $8.99$ \\ \hline 
 $ 1 $  & $44$ & $0.276$& $1.897$&$-0.318$  &$504$  &  $8.09$  \\ \hline 
\end{tabular}
\label{fittable}
\end{table}


\section{MCMC analysis} 
\label{sec:mcmc}

\begin{figure}[H]
\centering
  \includegraphics[width=18cm,height=10cm,keepaspectratio]{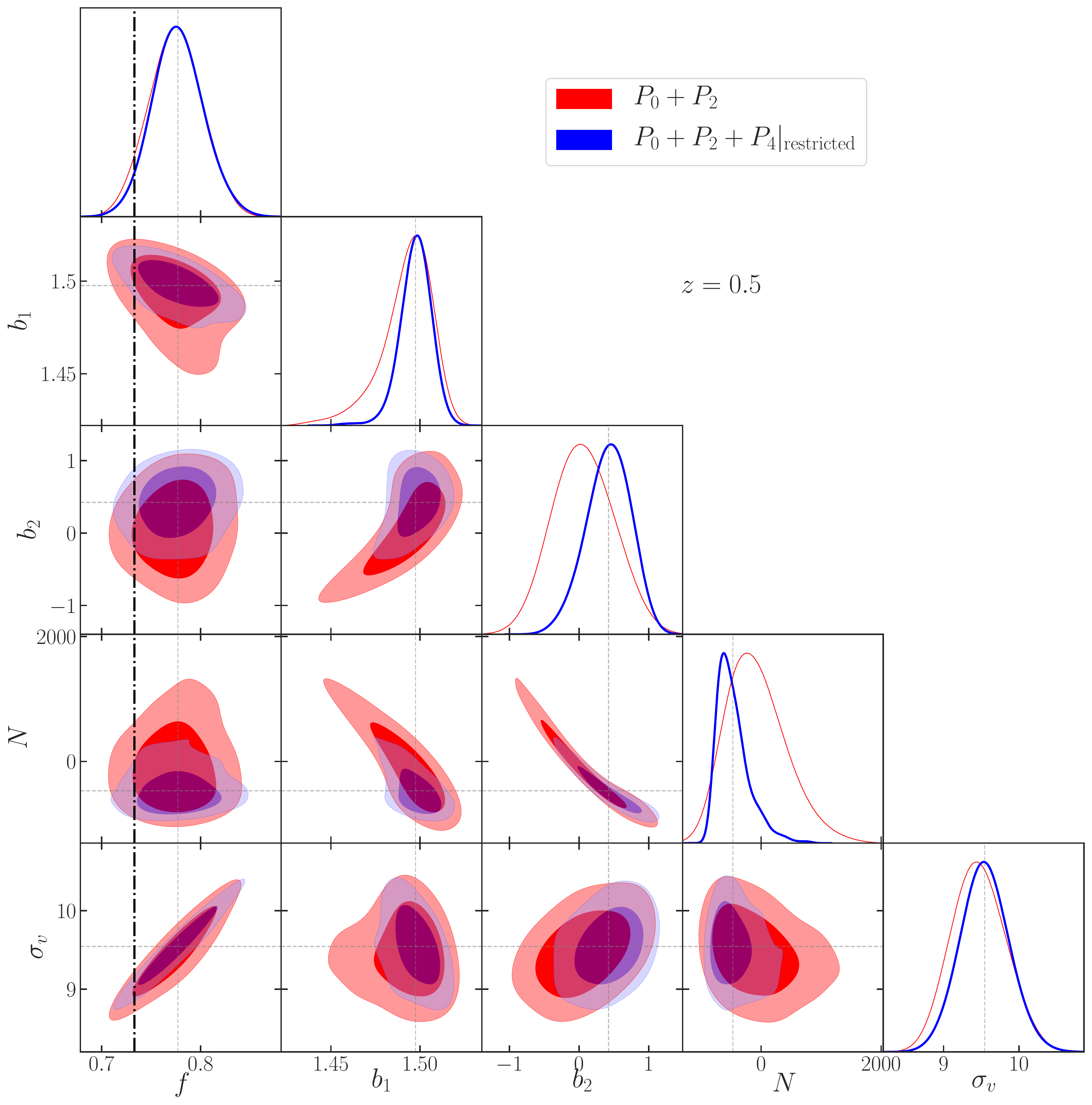}
  \includegraphics[width=18cm,height=10cm,keepaspectratio]{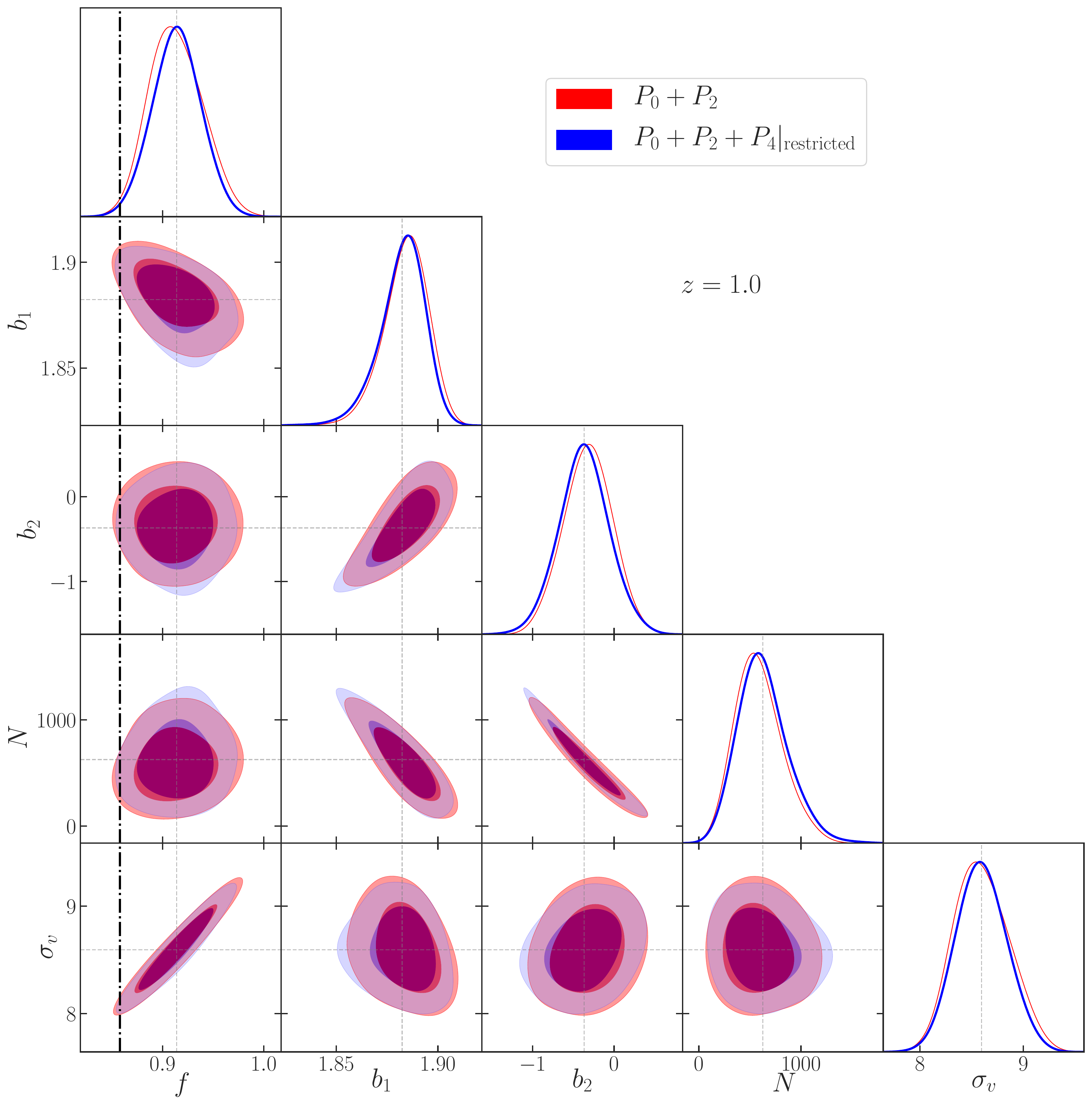}
  \caption[]{{\it Top:} MCMC results for the TNS model of \autoref{redshiftps} at $z=0.5$ with $k_{\rm max}=0.227 \, h$/Mpc for $P_0$ and $P_2$ and $k_{\rm max,4}=0.129 \, h$/Mpc for $P_4$. {\it Bottom:}  MCMC results for the TNS model of \autoref{redshiftps} at $z=1$ with $k_{\rm max}=0.276 \, h$/Mpc for $P_0$ and $P_2$ and $k_{\rm max,4}=0.05 \, h$/Mpc for $P_4$. In both panels, the thick black dashed-dot line indicates the fiducial value of $f$ in the PICOLA simulations. The thin grey dashed lines show the mean MCMC estimates for all parameters.}
\label{mcmc1}
\end{figure}

\begin{figure}[t!]
\centering
  \includegraphics[scale=0.6]{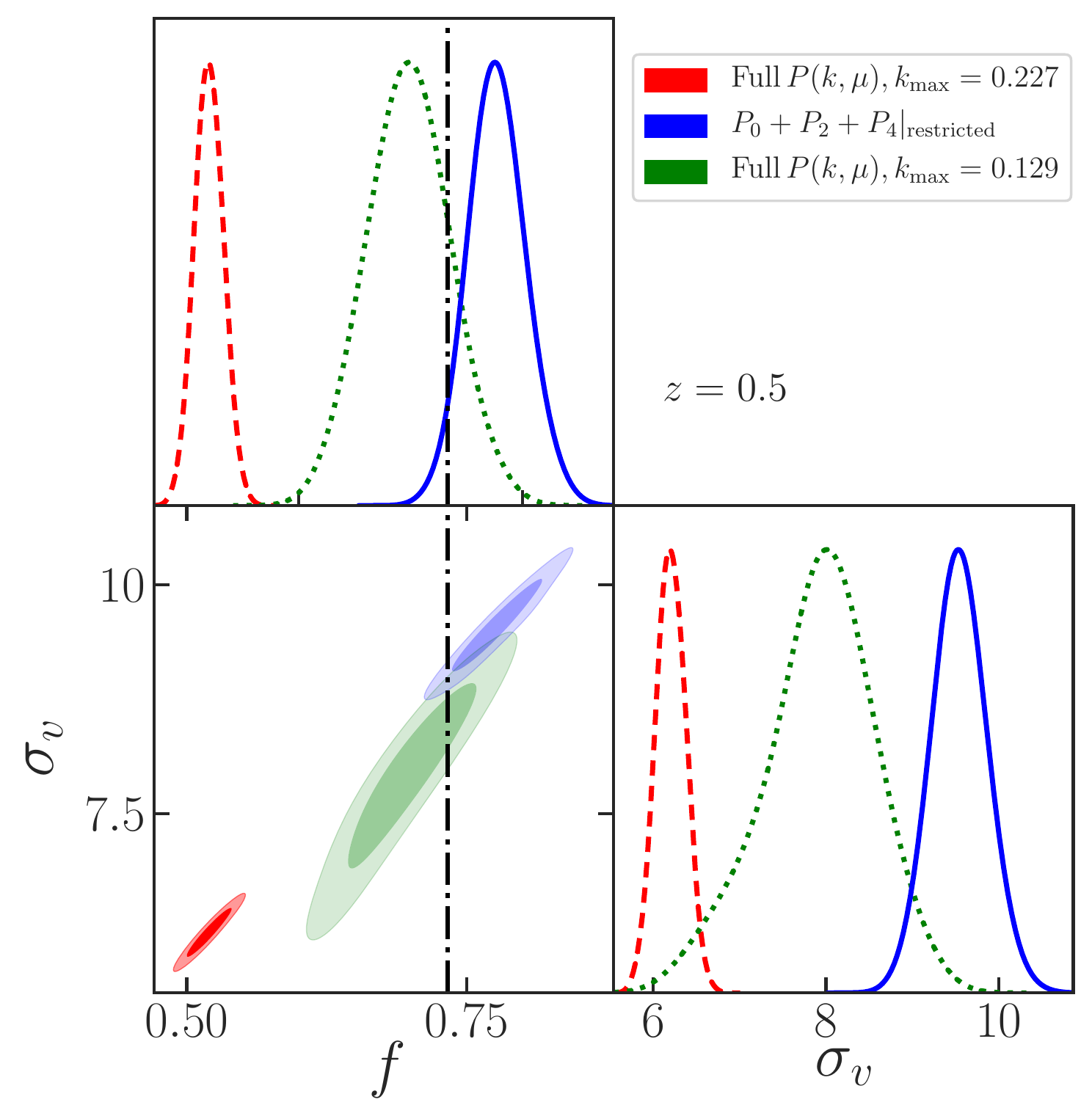}  
  \caption[]{MCMC comparison at $z=0.5$. We show the $P_0+P_2+P_4|_{\rm restricted}$ case as well as the full $P(k,\mu)$ case for two different values of $k_{\rm max}$, as detailed in the main text. The thick black dashed-dot line indicates the fiducial value of $f$ in the PICOLA simulations. The plot demonstrates the tradeoff between bias and constraining power in $f$ depending on the choice of $k_{\rm max}$.}
\label{fig:comp_bias}
\end{figure}

 In this section we perform Bayesian MCMC analyses at redshifts $z=0.5$ and $1$. We model our log-likelihood using \autoref{covarianceeqn}, and vary the growth rate $f$ and nuisance parameters of the TNS model outlined in \autoref{sec:model}.  
 \newline
 \newline
 Our approach has two purposes. The first is to check how biased the $f$ estimates can be at our derived $k_{\rm max}$ chosen in \autoref{fittable}. The second is to provide estimates for the $f$ constraints using our version of the TNS model and Stage IV-like specifications, and assess the improvement when adding the hexadecapole. 
\newline
\newline
Our results at $z=0.5$ are shown in the top panel of \autoref{mcmc1}. We first utilise the monopole and quadrupole spectra ($P_0$,$P_2$) only at the range of scales determined in \autoref{sec:model}.
The Figure shows the TNS model's recovery of $f$ at $k_{\rm max} = 0.227 \, h/$Mpc (red contours). We can see that the fiducial value is safely recovered with a $2\sigma$ criterion (same as in \autoref{sec:model}), and we have also checked that using a higher $k_{\rm max}$ the estimates become more biased.
\newline
\newline
We then add the hexadecapole, $P_4$. We find that the estimates of the growth rate $f$ is biased if we take the hexadecapole up to the same $k_{\rm max}$ found in \autoref{fittable}. That is because the TNS model is not flexible enough to account for the hexadecapole at this range of scales. This has also been seen in the BOSS data analysis in \citet{Beutler:2016arn}. However, motivated by \autoref{fits}, we can consider the hexadecapole up to a more conservative value, $k_{\rm max,4}=0.129 \, h/$Mpc, without biasing the $f$ estimate (blue contours). This is again similar to what has been done in the BOSS data analysis in \citet{Beutler:2016arn}. Our MCMC estimates for the fractional error on $f(z=0.5)$ with this process are $3.6\%$ using the monopole and quadrupole, and $3.2\%$ when adding the hexadecapole with the restricted range of scales. We will refer to this case as $P_0+P_2+P_4|_{\rm restricted}$ throughout the paper.
\newline
\newline
Our results at $z=1$ are shown in the bottom panel of \autoref{mcmc1}. We follow a similar procedure as before, with $k_{\rm max} = 0.276 \, h/$Mpc for the monopole and quadrupole, and $k_{\rm max,4}=0.05 \, h/$Mpc for the hexadecapole in order for the $f$ estimate to remain unbiased at our required $2\sigma$ level. We have checked that a higher $k_{\rm max,4}$ biases estimates of $f$. This $k_{\rm max,4}$ is lower than the one used at $z=0.5$ which may seem counterintuitive. We can explain this in terms of the model's flexibility. At $k_{\rm max} =0.276 \, h/{\rm Mpc}$ the model is already being severely tested and therefore it cannot account for $P_4$ to any higher $k_{\rm max,4}$ without becoming biased. Our MCMC estimates for the marginalised fractional error on $f(z=1)$ with this process are $3\%$ using the monopole and quadrupole, and $2.6\%$ when adding the hexadecapole with the restricted range of scales, $P_0+P_2+P_4|_{\rm restricted}$.
\newpage
Before moving on to the Fisher matrix analysis, it is useful to perform another set of MCMC analyses to get a sense of the tradeoff between bias and constraining power in $f$ depending on $k_{\rm max}$. At redshift $z=0.5$, in addition to the $P_0+P_2+P_4|_{\rm restricted}$ presented above we perform two additional MCMC analyses using the full $P(k,\mu)$ (equivalent to $P_0+P_2+P_4$ with equal $k_{\rm max}$ for monopole, quadrupole, and hexadecapole). We first set $k_{\rm max} =0.227 \, h/{\rm Mpc}$, followed by a more conservative $k_{\rm max} =0.129 \, h/{\rm Mpc}$. The results are shown in \autoref{fig:comp_bias}. For the full $P(k,\mu)$ with $k_{\rm max} =0.227 \, h/{\rm Mpc}$ (red dashed line) we find $f=0.520 \pm 0.013$, a heavily biased estimate compared to the true value of $f=0.733$ in the PICOLA simulations (thick black dashed-dot line). For the more conservative full $P(k,\mu)$ with $k_{\rm max} =0.129 \, h/{\rm Mpc}$ (green dotted line) we see that the true $f$ is within the $1\sigma$ contour, $f=0.698 \pm 0.037$. For the $P_0+P_2+P_4|_{\rm restricted}$ case, the estimate is unbiased at the $2\sigma$ level, $f=0.782 \pm 0.025$.

\section{Fisher matrix analysis}
\label{sec:fisher}

In this section we are going to perform an exploratory Fisher matrix analysis to forecast constraints on $f$ using the TNS model. Since we wish to compare the Fisher results with the ones from our MCMC analysis, we will use the Fisher matrix formalism written in terms of multipoles. This allows us to choose different ranges for the monopole, quadrupole, and hexadecapole spectra, to mimic the procedure followed in our MCMC analysis in the previous section. We will also perform a comparison between this method with the most commonly used full anisotropic power spectrum  method. This allows us to quickly assess how the requirement for unbiased $f$ estimates limits the improvement with respect to the case where the hexadecapole is added assuming the same $k_{\rm max}$ as in the monopole and quadrupole analysis.\\

\begin{figure}[t!]
\centering
  \includegraphics[width=25cm,height=9cm,keepaspectratio]{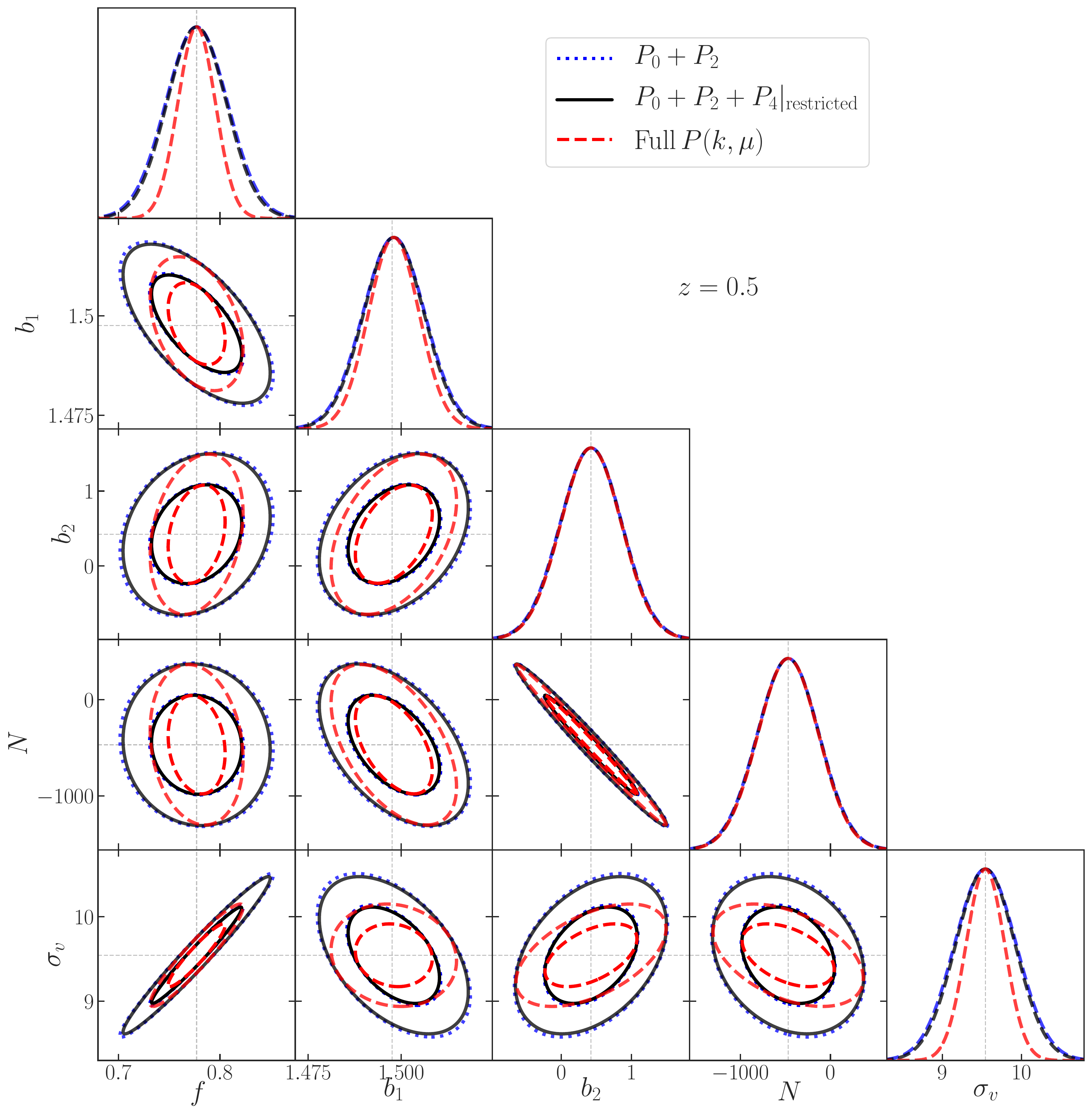}
  \includegraphics[width=25cm,height=9cm,keepaspectratio]{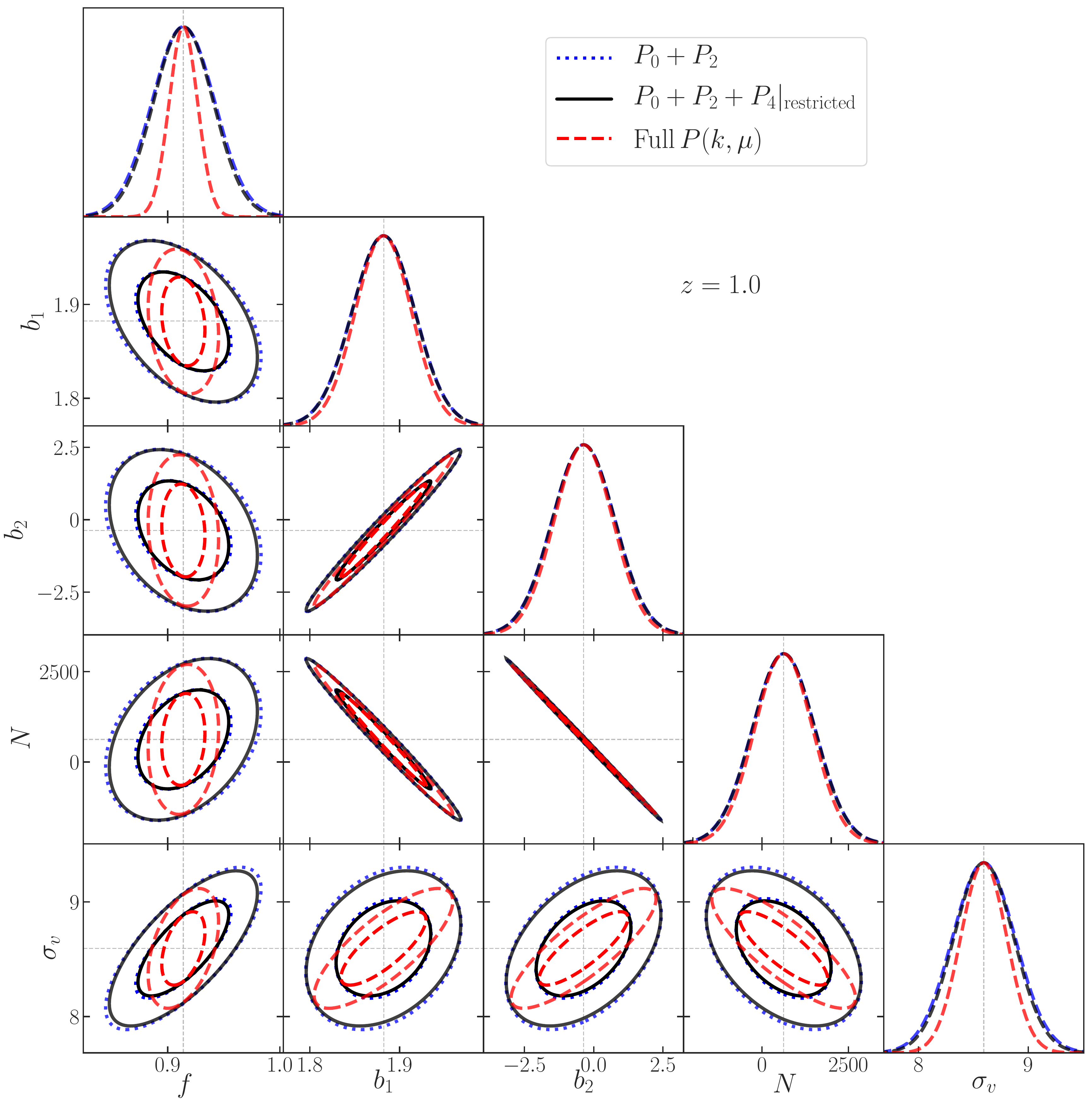}
  \caption[]{Fisher matrix forecasts for the TNS model of \autoref{redshiftps} at $z=0.5$ and $1$. The full $P(k,\mu)$ case is equivalent to taking $P_0$, $P_2$ and $P_4$ to the same $k_{\rm max}$. We also show $P_0+P_2$ only, as well as the $P_0+P_2+P_4|_{\rm restricted}$ case. Note that we have used the MCMC means as the fiducial Fisher matrix parameter values, to allow for a direct comparison with the MCMC.}
\label{fig:fisher1}
\end{figure}

\noindent
\emph{Fisher matrix formalism using multipole expansion}:\\ \\
\noindent
Here we summarise the multipole expansion formalism for Fisher forecasts. We refer the reader to \citet{Taruya:2011tz} for a comprehensive description. The authors of \citet{Taruya:2011tz} use the TNS model equiped with a linear deterministic bias, noting that this assumption is idealistic and a more accurate prescription is essential for realistic forecasts. They then perform a Fisher matrix analysis to investigate the relative contributions of the different multipoles ($P_0,P_2$ and $P_4$) if taken at the same $k_{\rm max}$, and compare with the full $P(k,\mu)$ formalism. Hence, an important distinction between the forecasts performed in this work and that of \citet{Taruya:2011tz} is (i) our implementation of a non-linear bias prescription that corresponds to a more accurate theoretical template that could readily be applied to the data (ii) the use of different $k_{\rm max}$ for the different multipoles to ensure unbiased forecasts and (iii) an MCMC analysis alongside the Fisher matrix analysis to consolidate our findings.
\newline
\newline
In terms of multipoles, the Fisher matrix for a set of parameters $\{p\}$ is given by \citep{Fisher:1935,Tegmark:1997rp, Seo:2007ns,Taruya:2011tz}
\begin{equation}
F_{ij}=\frac{V_{\rm s}}{4 \pi^{2}} \sum_{\ell, \ell^{\prime}} \int_{k_{\mathrm{min}}}^{k_{\mathrm{max}}} dk k^{2} \frac{\partial P^{\rm S}_{\ell}(k)}{\partial p_{i}}\left[\widetilde{\mathrm{Cov}}^{\ell \ell^{\prime}}(k)\right]^{-1} \frac{\partial P^{\rm S}_{\ell}(k)}{\partial p_{j}} \, ,
\label{eq:elFish}
\end{equation}
with $\widetilde{\mathrm{Cov}}^{\ell \ell^{\prime}}$ being the reduced covariance matrix:
\begin{equation}
\widetilde{\mathrm{Cov}}^{\ell \ell^{\prime}} (k)=\frac{(2 \ell+1)\left(2 \ell^{\prime}+1\right)}{2} \times \int_{-1}^{1} d \mu \mathcal{P}^{\rm S}_{\ell}(\mu) \mathcal{P}^{\rm S}_{\ell^{\prime}}(\mu)\left[P(k, \mu)+\frac{1}{\overline{n}}\right]^{2} \, .
\label{eq:cov}
\end{equation}
Note that in our analysis, we promote $k_{\rm max} \rightarrow k_{{\rm max},\ell}$ in \autoref{eq:elFish}, in order to be able to study the $P_0+P_2+P_4|_{\rm restricted}$ case. Also, since we want to mimic our assumptions in the MCMC analysis, we use linear covariance, which means that $P(k,\mu)$ in the brackets of \autoref{eq:cov} is given by the linear formula \citep{Kaiser:1987qv}. \\ \\
\noindent
\emph{Fisher matrix formalism using the full (2D) anisotropic power spectrum}: \\ \\
\noindent
Considering the full power spectrum signal in redshift space, the Fisher matrix becomes \citep{Tegmark:1997rp, Seo:2007ns}
\begin{equation}
F_{i j}^{(2 \mathrm{D})}=\frac{V_{\rm s}}{4 \pi^{2}} \int_{-1}^{1} d \mu \int_{k_{\min }}^{k_{\max }} d k k^{2} \frac{\partial P(k, \mu)^{\rm S}}{\partial p_{i}}\left\{P(k, \mu)+\frac{1}{\overline{n}}\right\}^{-2}\frac{\partial P(k, \mu)^{\rm S}}{\partial p_{j}} \, .
\label{eq:Fish}    
\end{equation}
Using the Fisher matrix formalism, the forecasted errors on parameter $p_i$, marginalised over all other parameters, are given by the square root of the diagonal of the inverse of the Fisher matrix as
\begin{equation}
\Delta p_i = \sqrt{(F^{-1})_{ii}} \, . 
\end{equation}
We are now ready to present our Fisher matrix analysis results. In \autoref{fig:fisher1} we compare 
the full $P(k,\mu)$ Fisher matrix results\footnote{We have checked that this is equivalent to using the multipole expansion formalism taking $P_0$, $P_2$, and $P_4$ up to the same $k_{\rm max}$ \citep{Taruya:2011tz}. This is expected since higher order multipoles carry little information \citep{Hand:2017ilm}.} with the results using only $P_0$ and $P_2$ up to the $k_{\rm max}$ determined in \autoref{fittable}, and the ones found when adding the hexadecapole with the restricted range of scales in order for the $f$ estimates to remain unbiased ($P_0+P_2+P_4|_{\rm restricted}$).
\newline
\newline
In both redshifts, $z=0.5$ and $z=1$, it is clear that the full $P(k,\mu)$ treatment gives much better constraints on $f$. More specifically, using the full $P(k,\mu)$ formalism we find a $2.4\%$ constraint on $f$ compared to $3.8\%$ in the $P_0+P_2+P_4|_{\rm restricted}$ case at $z=0.5$. At $z=1$, we find $1.4\%$ compared to $2.9\%$ in the $P_0+P_2+P_4|_{\rm restricted}$ case, a factor of $\sim 2$ difference. 

\subsection{Comparison between Fisher matrix and MCMC analysis}

\begin{figure}[t!]
\centering
  \includegraphics[scale=0.4]{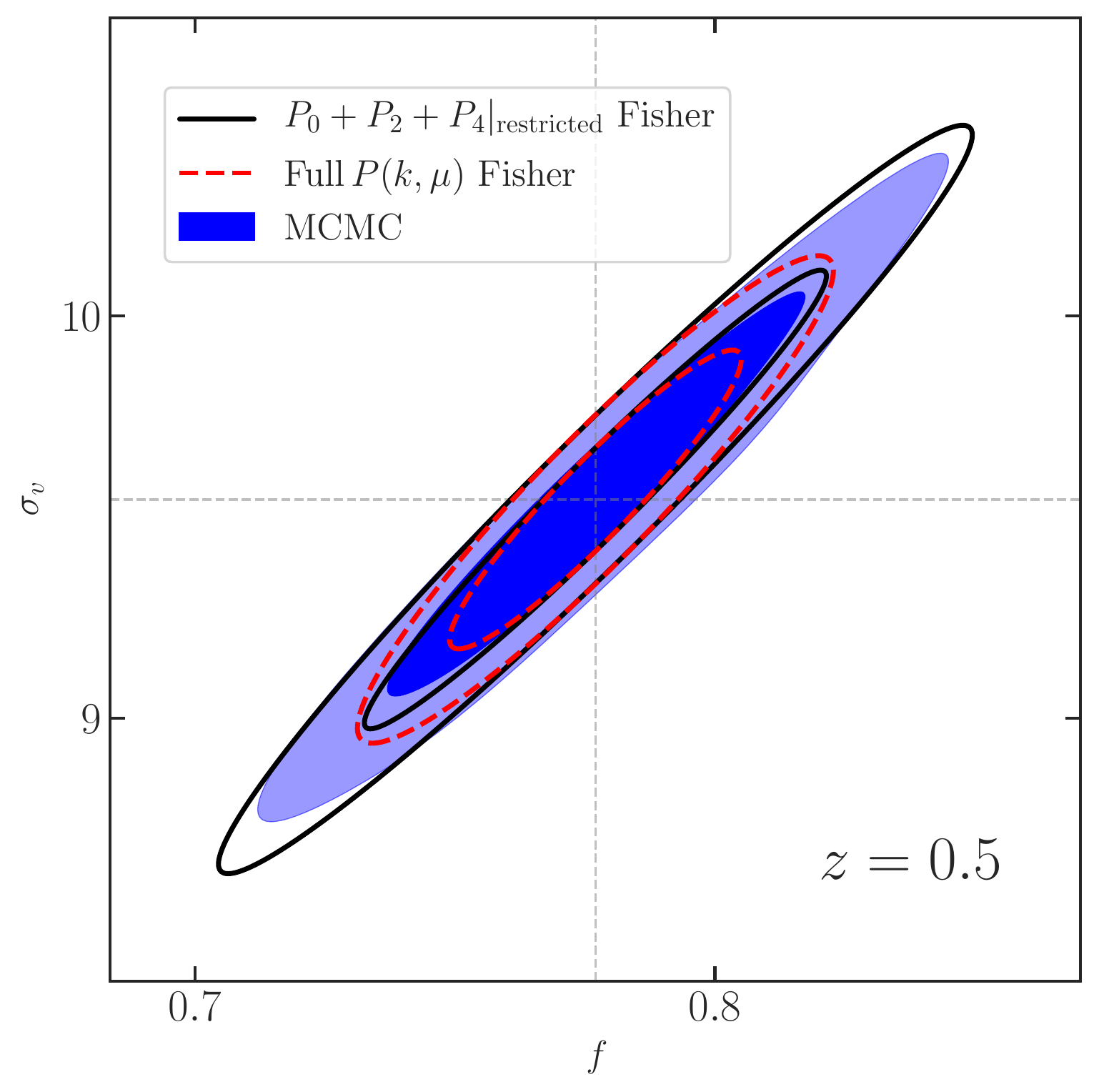}  
   \includegraphics[scale=0.4]{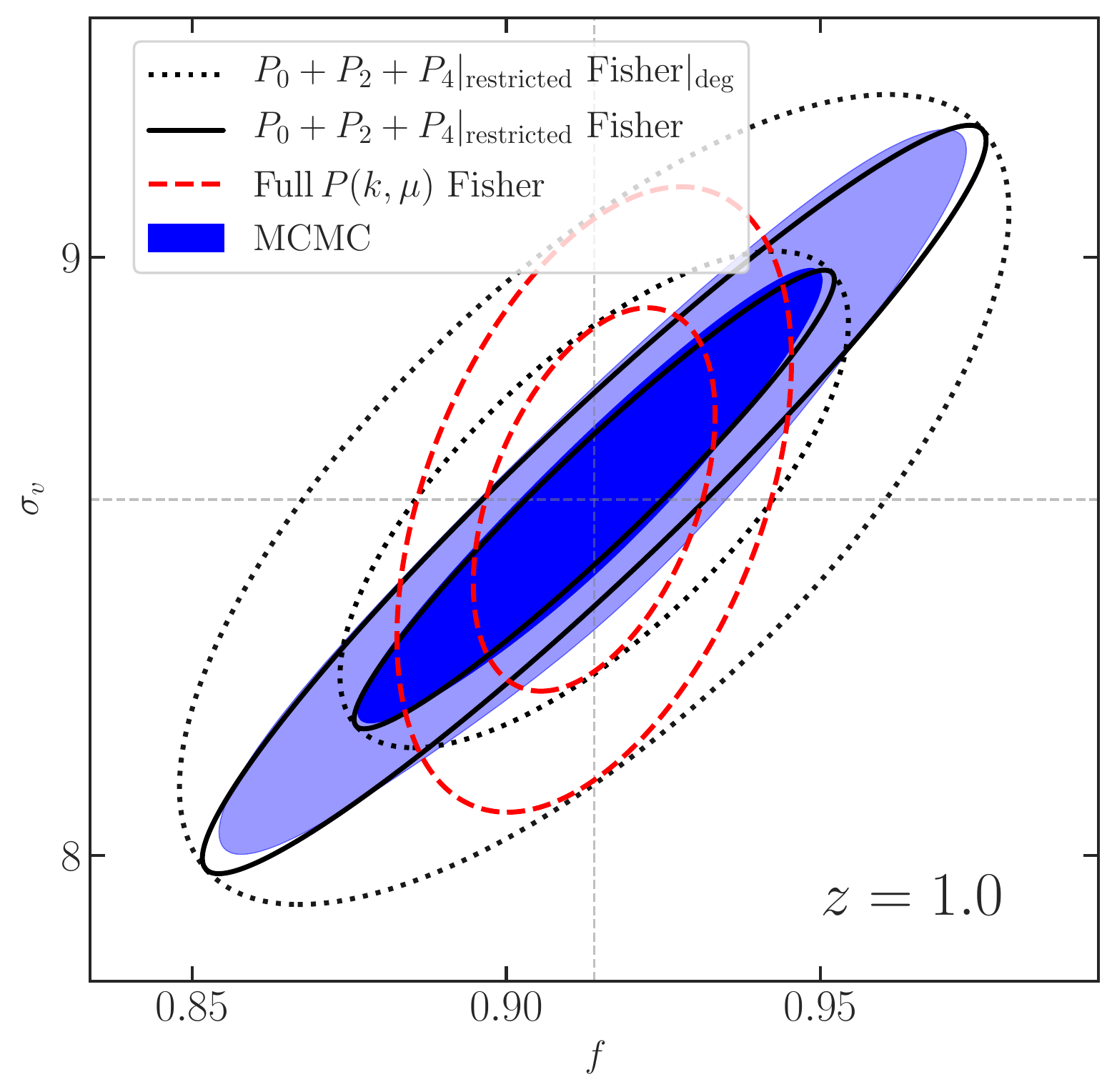}
  \caption[]{Fisher and MCMC comparison for the $P_0+P_2+P_4|_{\rm restricted}$ case. We also show the full $P(k,\mu)$ Fisher ellipses, but note these correspond to biased $f$ estimates (see \autoref{fig:comp_bias}). As described in the main text, the ${\rm Fisher}|_{\rm deg}$ contour (dotted line) at $z=1$ corresponds to the Fisher matrix shown in the bottom panel of \autoref{fig:fisher1}, with a near perfect degeneracy between the $(b_2,N)$ nuisance parameters. The ${\rm Fisher}$ contour (solid line) is the result after imposing a conservative prior on $N$ that breaks the degeneracy just enough to mitigate the instability it induces. We emphasise that in both cases the final marginalised $f$ constraints remain stable and in very good agreement with the MCMC.}
\label{fig:comp1}
\end{figure}

We will now present the comparison between Fisher matrix and MCMC results, in the unbiased, $P_0+P_2+P_4|_{\rm restricted}$ case. For this purpose we will focus on comparing the $(f,\sigma_v)$ contours, since $\sigma_v$ is the most correlated nuisance parameter with the cosmological parameter of interest $f$. We include the full range of comparison plots in \autoref{sec:appendix}. \\
\newline
The results are shown in \autoref{fig:comp1}. At $z=0.5$, we find remarkable agreement between Fisher and MCMC, demonstrating that the multipole expansion formalism method is both reliable and robust. At $z=1$, we have plotted two Fisher ellipses: the dotted one, denoted by $\mathrm{Fisher|_{deg}}$, corresponds to the Fisher matrix results presented in \autoref{fig:fisher1}. We see that the area of the $(f,\sigma_v)$ ellipse is larger than the MCMC one. 
\noindent
Inspecting \autoref{fig:fisher1} at $z=1$, we see that there is a near-perfect degeneracy between the $(b_2,N)$ parameters\footnote{A  contributing factor to this could be the loop integral involving $b^2_2$, that asymptotes to a constant at low $k$ \citep{Desjacques:2016bnm}.}. We can quantify this using the \emph{correlation coefficient} $r$ given by
\begin{equation}
r(p_i,p_j) = \frac{(F^{-1})_{ij}}{\sqrt{(F^{-1})_{ii}(F^{-1})_{jj}}} \, .
\end{equation}
This characterises parameter degeneracies: $r=0$ means $p_i$ and $p_j$ are uncorrelated, while $r = \pm 1$ means they are completely (anti)correlated. In the case we are concerned with at $z=1$, we find $r(b_2,N)=-0.999$.
This signals a possible instability in the Fisher matrix that might be responsible for the disagreement with the MCMC. To investigate this, we impose a conservative $\sim 50\%$ prior on the $N$ parameter and rerun the Fisher matrix analysis. This breaks the degeneracy just enough to mitigate the instability, and gives the excellent agreement shown with the solid black line. 
However, we  emphasise that despite this instability, the marginalised constraint on $f$ is in both cases (with and without the prior correction) in very good agreement with the MCMC. This is because it is the $(f,\sigma_v)$ degeneracy that affects mostly the final, marginalised error on $f$.
The careful reader will notice that there are some differences between the Fisher and MCMC results, especially regarding the improvement when including the hexadecapole at $z=0.5$. This seems to give only marginal gains in the constraining power when considering the Fisher matrix, disagreeing with the improvement shown by the MCMC forecast. Given that the MCMC contours are not perfect Gaussian ellipses we would not expect perfect agreement with the Fisher ellipses. The disagreement is worse at $z=0.5$ and the MCMC contours there also look less ``Gaussian'' than the ones at 
$z=1$, so this is not very surprising. 
We note that similar subtleties have been seen before in Fisher and MCMC comparison studies, and we refer the interested reader to \citet{Wolz:2012sr,Hawken:2011nd}. The final, marginalised $f$ constraints from all methods we have used are summarised in \autoref{constraintstab}.

\section{Summary and conclusions}
\label{sec:summary}

In this paper we have assessed the performance of the commonly used TNS model in the context of Stage IV galaxy surveys, considering the multipoles of the redshift space halo power spectrum as our observable. We considered two redshifts, $z=0.5$ and $z=1$, and made use of PICOLA simulations to perform maximum likelihood, MCMC, and Fisher matrix analyses. Here we summarise our main results and conclude.
\\ \\
We first determined the TNS model's ranges of validity using the monopole and quadrupole power spectra, which contain most of the cosmological information. We note that while this approach is appropriate for the purposes and limitations of this work, in an actual data analysis the bias on $f$ needs to be evaluated directly as a function of $k_{\rm max}$. In \cite{Bose:2019ywu}, this point is investigated thoroughly by running a large number of MCMC analyses. In the MCMC analysis that followed, we varied the model's four nuisance parameters and the logarithmic growth rate, $f$. 
\begin{table}[H]
\centering
\caption{Marginalised percent errors (at $1\sigma$) on $f$ from the MCMC and Fisher analyses at $z=0.5$ and $z=1$. We utilise the monopole and quadrupole up to the $k_{\rm max}$ given in \autoref{fittable}, and the hexadecapole up to a conservative (restricted) $k_{\rm max,4}$, as detailed in the main text. For comparison we show the results for the full $P(k,\mu)$ Fisher matrix case, and warn that this falsely small error corresponds to biased $f$ estimates.} 
\begin{tabular}{| c | c | c |}
\hline  
 Analysis & $z=0.5$ & $z=1$ \\
 \hline
 MCMC: $P_0+P_2$ & $3.6\%$ & $3.0\%$  \\ \hline 
 Fisher: $P_0+P_2$ & $3.9\%$ & $3.1\%$  \\ \hline
 MCMC: $P_0+P_2+P_4|_{\mathrm{restricted}}$ & $3.2\%$ & $2.6\%$   \\ \hline 
 Fisher: $P_0+P_2+P_4|_{\mathrm{restricted}}$ & $3.8\%$ & $2.9\%$   \\ \hline \hline
 Fisher: Full $P(k,\mu)$, biased & $2.4\%$ & $1.4\%$   \\ \hline
\end{tabular}
\label{constraintstab}
\end{table}
\noindent
The analysis using the $k_{\rm max}$ from \autoref{fittable} shows a significant degeneracy between $f$ and $\sigma_v$ that has also been found previously \citep{Zheng:2016xvo,Bose:2017myh}. The improvement on the TNS constraints at $z=1$ is mainly due to the much higher $k_{\rm max}$ at $z=1$ compared to that at $z=0.5$. 
\\ \\
To investigate the possible improvement by adding information from the hexadecapole, we 
performed two distinct MCMC analyses at $z=0.5$ and $1$. One excludes the hexadecapole using the range of validity found in \autoref{sec:model}, and one includes it. It has been shown that adding the hexadecapole up to the same $k_{\rm max}$ as the monopole and quadrupole biases the estimation of $f$ (see also \autoref{fig:comp_bias}). We followed what was done in the BOSS analysis in \citet{Beutler:2016arn}, and restricted its range to a maximum $k_{\rm max,4}=0.129 \, h/{\rm Mpc}$ for $z=0.5$ and $k_{\rm max,4}=0.05 \, h/{\rm Mpc}$ for $z=1$, so that the estimation of $f$ remains unbiased at the required $2\sigma$ level. Our results are summarised in \autoref{constraintstab}. 
\\ \\
The addition of the hexadecapole with a restricted $k_{\rm max}$ improves the $f$ constraints, but at a much lower level than using the full $P(k,\mu)$ method, which is equivalent to taking $P_0, P_2$ and $P_4$ up to the same $k_{\rm max}$ and leads to biased estimates of $f$. 
\\ \\
Finally, we performed a comprehensive Fisher matrix analysis to quickly explore the parameter space and test whether we can reproduce our MCMC analysis results. Using the multipole expansion formalism for the Fisher matrix we reached very good agreement with the MCMC, as shown in \autoref{constraintstab}. We also compared the $P_0+P_2+P_4|_{\rm restricted}$ case with the full $P(k,\mu)$ case, showing that the former gives much more conservative error estimates (as well as avoiding bias in the estimate of $f$ itself). 
\\ \\ 
It is useful to try and assess how our forecasted constraints on the growth rate, given the $k_{\rm max}$ cuts, compare to current requirements for Stage IV surveys. We will refer to the Euclid-like forecasts for the growth rate $f$ presented in \citet{Amendola:2016saw} (see Table 4). At $z=1$ with the same survey volume as we use, they find a fractional error on f of $1.4\%$ in their \emph{pessimistic} scenario, which has assumed a number density $16\%$ lower than ours. Their adopted model is linear Kaiser multiplied by the BAO smearing function proposed in \citep{Eisenstein:2006nj}. They use the full $P(k,\mu)$ with a $k_{\rm max} \sim 0.2 \, {\rm h/Mpc}$. Their set of redshift dependent parameters includes the growth rate $f(z)$, the linear bias $b(z)$, the residual shot noise $P_{\rm s}$, the angular diameter distance $D_A(z)$, and the Hubble rate $H(z)$, but the latter two are not marginalised over. Instead, they are projected onto the set of parameters they depend on, which helps with the information on the background (shape) parameters they also vary. Since the models, parameter sets, forecasting method, and $k_{\rm max}$ choices are different, it is very difficult to compare directly, but note that we find the same fractional error of $1.4\%$ using the full $P(k,\mu)$, and a factor of 2 larger error using the hexadecapole cut needed to keep $f(z)$ unbiased at the $2\sigma$ level.
\\ \\
In summary, our study reinforces the need for accurate modelling of the non-linear redshift space power spectrum in light of upcoming Stage IV galaxy surveys. It also shows how reliable forecasts can be obtained if the forecast procedure closely follows what is being done in real data analyses. To our knowledge, a forecasting galaxy clustering analysis for Stage-IV surveys using an accurate theoretical template that can be readily applied to data, and the multipole expansion formalism with different, realistic $k_{\rm max}$ limits informed by simulations, has not been performed in the literature. In addition, we stress that even if a fit to simulations is not available, one should still perform conservative forecasts using the multipole expansion formalism with different ranges for the different multipoles. 
\\
\\
That being said, the model considered here is partly phenomenological, and ideally one would want a fully perturbative model that consistently models bias and redshift space distortions, and under which one can obtain a handle on the modelling uncertainties in very general cosmological settings. Several studies have been performed in this direction recently \citep[see, for example,][]{Desjacques:2018pfv,Desjacques:2016bnm,Ivanov:2018gjr,Ding:2017gad,Hand:2017ilm}.
In the next paper of this series \citep{Bose:2019psj} we also push in this direction and perform an extensive investigation using other proposed models like the Effective Theory of Large Scale Structure \citep[see, for example,][]{Perko:2016puo,delaBella:2018fdb,Lewandowski:2015ziq}, and compare their performance to the TNS model.

\section*{Acknowledgments}
\noindent We would like to thank the two anonymous referees for their very useful comments and suggestions that improved the quality of this manuscript. We are grateful to Hans Winther for providing the simulation data. We thank Florian Beutler for useful discussions. We acknowledge use of open source software \citep{scipy:2001,Hunter:2007, Lewis:1999bs, mckinney-proc-scipy-2010, numpy:2011}. KM acknowledges support from the UK Science \& Technology Facilities Council through grant ST/N000668/1 and from the UK Space Agency through grant ST/N00180X/1. BB acknowledges support from the Swiss National Science Foundation (SNSF) Professorship grant No.170547. AP is a UK Research and Innovation Future Leaders Fellow, grant MR/S016066/1, and also acknowledges support from the UK Science \& Technology Facilities Council through grant ST/S000437/1.
Numerical computations for this research were done on the Sciama High Performance Compute (HPC) cluster which is supported by the ICG, SEPNet, and the University of Portsmouth.

\appendix
\section{Fisher and MCMC comparison for the $P_0+P_2+P_4|_{\rm restricted}$ case}
\label{sec:appendix}

\begin{figure}[H]
\centering
  \includegraphics[width=18cm,height=8cm,keepaspectratio]{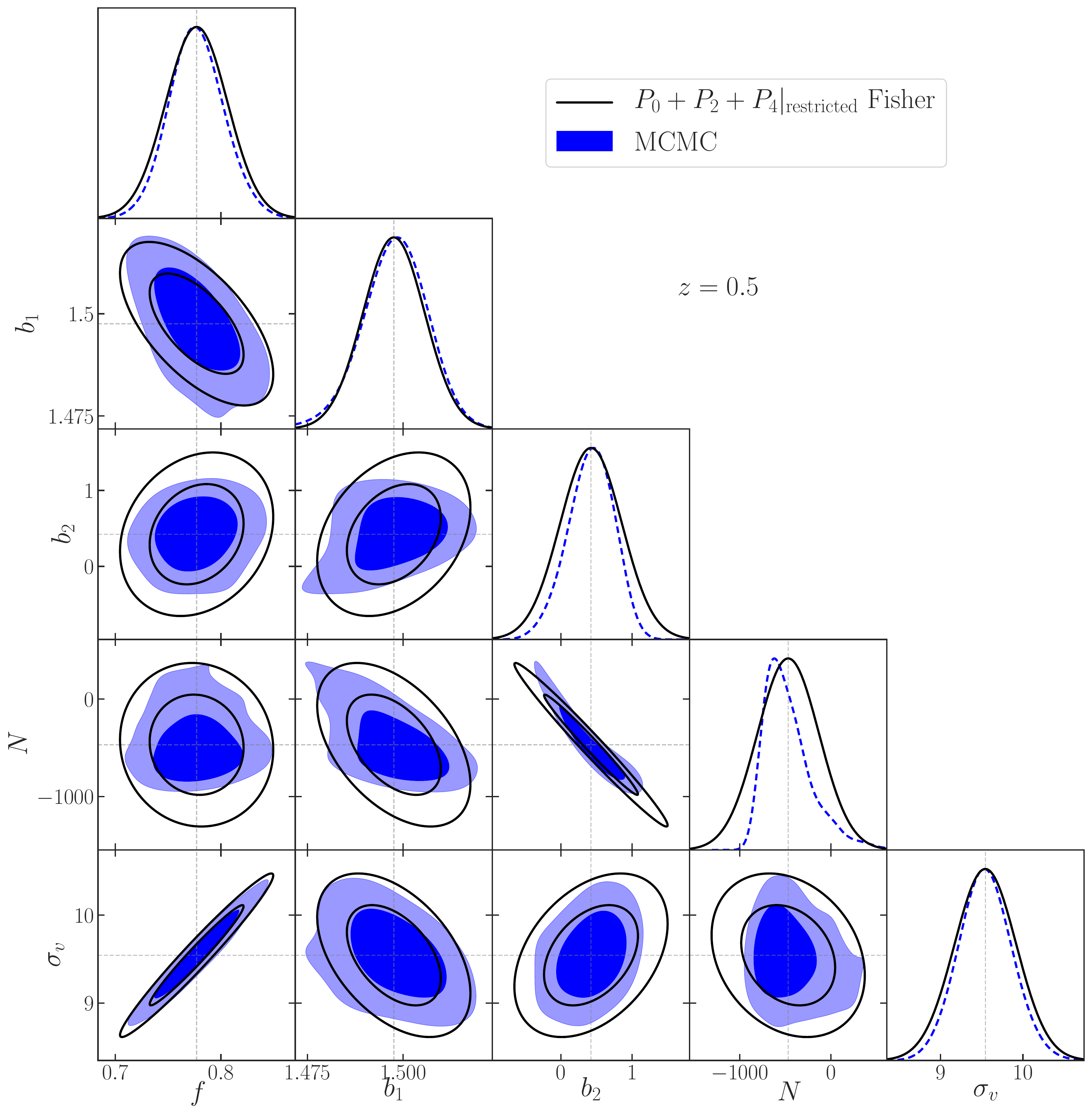}
  \includegraphics[width=18cm,height=8cm,keepaspectratio]{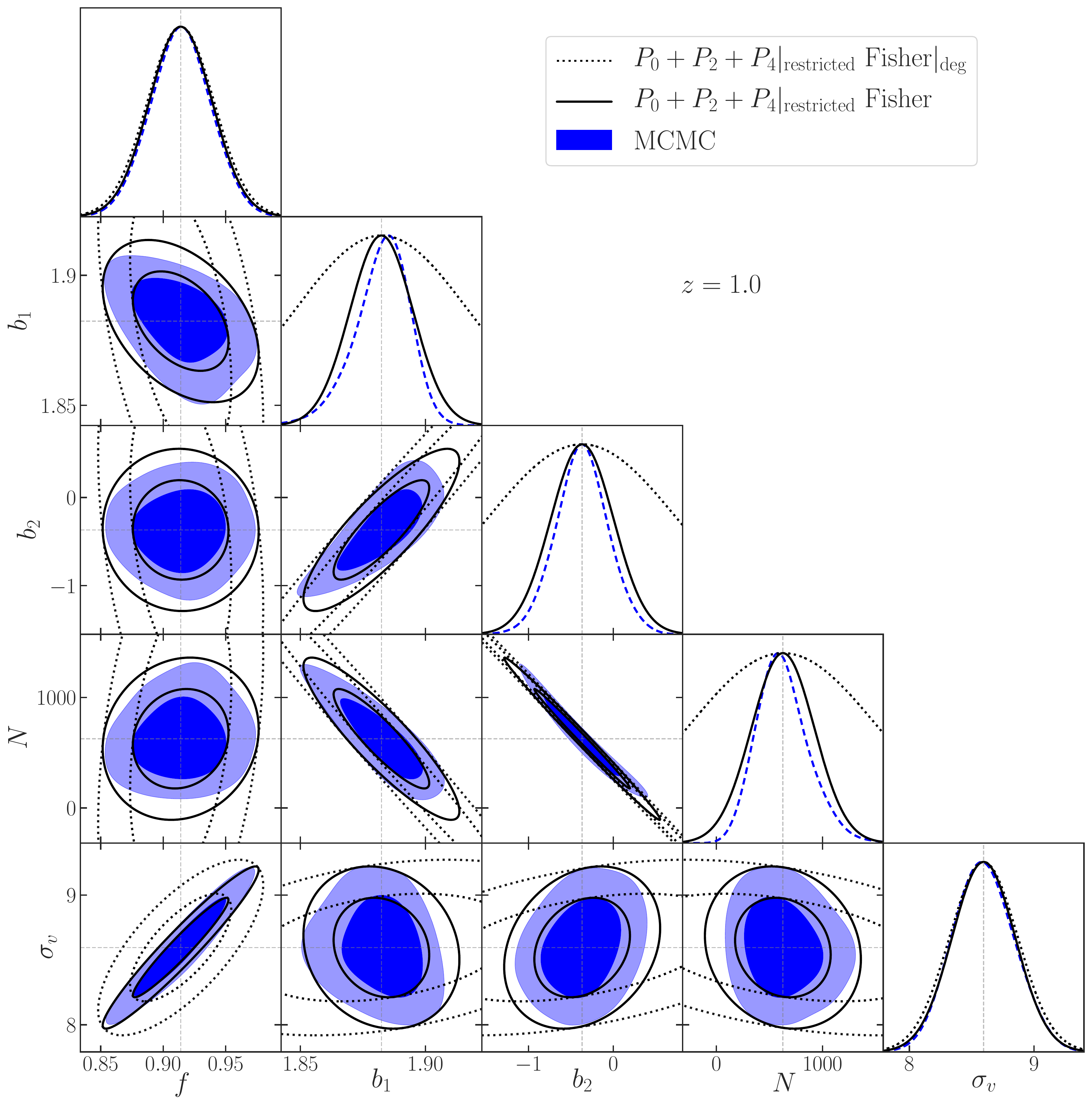}
  \caption[]{Complete Fisher and MCMC comparison plots for the $P_0+P_2+P_4|_{\rm restricted}$ case. Despite the deviations from the Gaussian shape in some of the MCMC contours, the final marginalised constraints on the cosmological parameter of interest $f$ are in  agreement.}
\label{mcmc1full}
\end{figure}

\bibliography{mybib} 
\bibliographystyle{apsrev4-1}

\end{document}